\documentclass[preprint,showpacs,preprintnumbers,amsmath,amssymb]{revtex4}

\usepackage{graphicx}
\usepackage{dcolumn}
\usepackage{bm}
\usepackage{booktabs}

\begin{document}

\title{\large Proximity effects and pair currents in cuprate junctions }

\author{Gad Koren}
\email{gkoren@physics.technion.ac.il} \affiliation{Department of Physics,
Technion - Israel Institute of Technology, Haifa
32000, ISRAEL} \homepage{http://physics.technion.ac.il/~gkoren}


\date{\today}
\def\bfig {\begin{figure}[tbhp] \centering}
\def\efig {\end{figure}}

\normalsize \baselineskip=8mm  \vspace{15mm}


\begin{abstract}

Proximity effects and pair currents were measured in epitaxial trilayer \textit{c-axis} junctions comprise of a $PrBa_2Cu_3O_{7-\delta}$ barrier sandwiched in between an overdoped $Y_{0.94}Ca_{0.06}Ba_2Cu_3O_{7-\delta}$ and underdoped $YBa_2Cu_{2.7}Co_{0.3}O_y$ layers. These junctions had two $\rm T_c$ values of $\rm T_c(high)=84-86$ K and $\rm T_c(low)=50-55$ K, allowing investigation when both electrodes are superconducting, or when only one is superconducting while the other is in its pseudogap regime. For T below $\rm T_c(high)$ but much above $\rm T_c(low)$, two distinct proximity effect transitions were observed in the resistance at two temperature regimes, between 80 and 84 K, and 76 to 80 K. The first is a conventional proximity effect with the $\rm T_c(high)$ electrode, while the second is a second-order proximity effect of this electrode with uncorrelated pairs in the pseudogap regime. Conductance spectra measured between 2 and 86 K showed four different $I_c$ pair currents which were attributed to coherent pairs tunneling through the barrier below 42 K, to fluctuating pairs current up to $\sim$77 K, and to proximity pairs current between 77 and 84 K. All pair currents were suppressed under magnetic fields, with two distinct decay parameters that originated in the two different electrodes, with a significant suppression observed in the pseudogap regime.

\end{abstract}

\maketitle

\section{Introduction}

Almost three decades after the discovery of the pseudogap in the cuprates \cite{Timusk-Statt,Norman} its origin is still unknown. Experimental results show that its commonly observed cross over temperature $T^*$ in many physical parameters, is actually due to a real phase transition of magnetic and elastic properties \cite{Fauque,Shekhter}. Other phenomena however, such as the Nernst effect and charge or pair density waves (CDW and PDW) were found to occur at temperature much lower than $T^*$, but still in the pseudogap regime \cite{Ong,Keimer}. Various models were proposed to account for the pseudogap phase ranging from fluctuating, uncorrelated pre-formed pairs which are precursors to superconductivity \cite{Kivelson}, to competing or coexisting orders with superconductivity \cite{KeimerRev}. Excess currents due to pairs fluctuation in the pseudogap state were observed in copper-oxide superconductors \cite{Bergeal}. Following a recently proposed  model of Amperian-pairing based on the competing PDW order \cite{Patrick}, we set up an experiment to test whether such pairing actually exists or not in planar \textit{c-axis} junctions of the cuprates under in-plane magnetic fields. These junctions comprised of a trilayer base electrode having a $PrBa_2Cu_3O_{7-\delta}$ (PrBCO) barrier sandwiched in between two $YBa_2Cu_3O_{7-\delta}$ (YBCO) or doped YBCO layers, and covered by a thick gold electrode. As no Amperian-pairing was found in this experiment, we investigated instead the fluctuating pair lifetimes and supercurrents in these junctions, and found two distinct lifetimes in the pseudogap regime, as well as a fluctuating pairs current much above $T_c$ of the underdoped electrode \cite{KorenLee}. In the present study we focus on the results observed in junctions of one particular wafer, which were not elaborated on in our previous study. This wafer which had an underdoped electrode, showed a new kind of a proximity effect transition, four pair currents of different origins, as well as strong suppression under magnetic field of the pairs fluctuation current  in the pseudogap regime. We found that the presence of the underdoped electrode with its pseudogap in these junctions was essential, as no such results were observed in junctions on other wafers which had no underdoped electrode \cite{KorenLee}. \\

\section{Experimental}

For our previous study \cite{KorenLee}, fifty \textit{c-axis} junctions were prepared on five different wafers, ten junctions of the same type on each wafer, as described in detail in Table I. The junctions structure and fabrication process are basically similar to that described previously \cite{Kirzhner}, and further elaborated on in Ref. \cite{KorenLee}, so there is no need to add more details here. All junctions had a low-$T_c$ electrode with $T_c\approx 50-55$ K [$T_c(low)$], and a high-$T_c$  electrode with $T_c\approx 85-90$ K [$T_c(high)$]. The idea was to measure mostly at temperatures in between these two transition temperatures, where fluctuations of the low-$T_c$ electrode could be probed by the order parameter of the high-$T_c$ electrode, thus enabling measurements of supercurrents and pair lifetimes in the junctions in the fluctuations regime, as discussed theoretically by Scalapino even before the cuprates were discovered \cite{Scalapino}. In the present study we focus on the transport results observed on junctions of the CJ-4 wafer, which had an overdoped $Y_{0.94}Ca_{0.06}Ba_2Cu_3O_{7-\delta}$ electrode and an underdoped YBCoCO electrode, marked as layers S1 and S2 in Fig. 1 (a), respectively. This figure shows a schematic cross-section of the trilayer base electrode of these junctions, while a typical Atomic Force Microscope image of one such junction including the gold cover electrode is given in Fig. 1 (b).  Transport measurements were carried out using the four-probe technique, with and without a magnetic field of up to 8 T, parallel or perpendicular to the wafer.\\

\begin{table}[h!]
  \centering
  \caption{\textit{c-axis} junction parameters. YBCO and PrBCO are optimally doped $YBa_2Cu_3O_{7-\delta}$ and $PrBa_2Cu_3O_{7-\delta}$, respectively and YBCoCO is underdoped $YBa_2Co_{0.3}Cu_{2.7}O_y$. All junctions were prepared on (100) $SrTiO_3$ wafers. Last column is the overlap junction area.}
  \label{tab:table1}
  \begin{tabular}{ccccc}
   \toprule[0.75pt] \toprule[0.75pt]
    wafer \# & layer 1 & layer 2 & layer 3 & area ($\mu m^2$)\\
    \midrule[0.5pt]
    CJ-1 \,\,& 300nm YBCO \,\,\,& 50nm PrBCO \,\,\,& 100nm YBCoCO\,\,\, & $ 7\times 5$\\
    CJ-2 \,\,& 200nm YBCO \,\,\,& 25nm PrBCO \,\,\,& 100nm YBCoCO\,\,\, & $20\times 15$\\
    CJ-4 \,\,& 200nm  $Y_{0.94}Ca_{0.06}Ba_2Cu_3O_y$\,\,\,& 25nm PrBCO \,\,\,& 100nm YBCoCO\,\,\, & $20\times 15$\\
    CJ-5 \,\,& 200nm YBCO \,\,\,& 25nm PrBCO \,\,\,& 100nm $Y_{0.7}Ca_{0.3}Ba_2Cu_3O_y$\,\,\, & $20\times 15$\\
    CJ-6 \,\,& 200nm $Y_{0.94}Ca_{0.06}Ba_2Cu_3O_y$ \,\,\,& 25nm PrBCO \,\,\,& 100nm $Y_{0.7}Ca_{0.3}Ba_2Cu_3O_y$\,\,\, & $7\times 5$\\
    \bottomrule[0.75pt]
  \end{tabular}
\end{table}

\begin{figure} \hspace{-20mm}
\includegraphics[height=6cm,width=14cm]{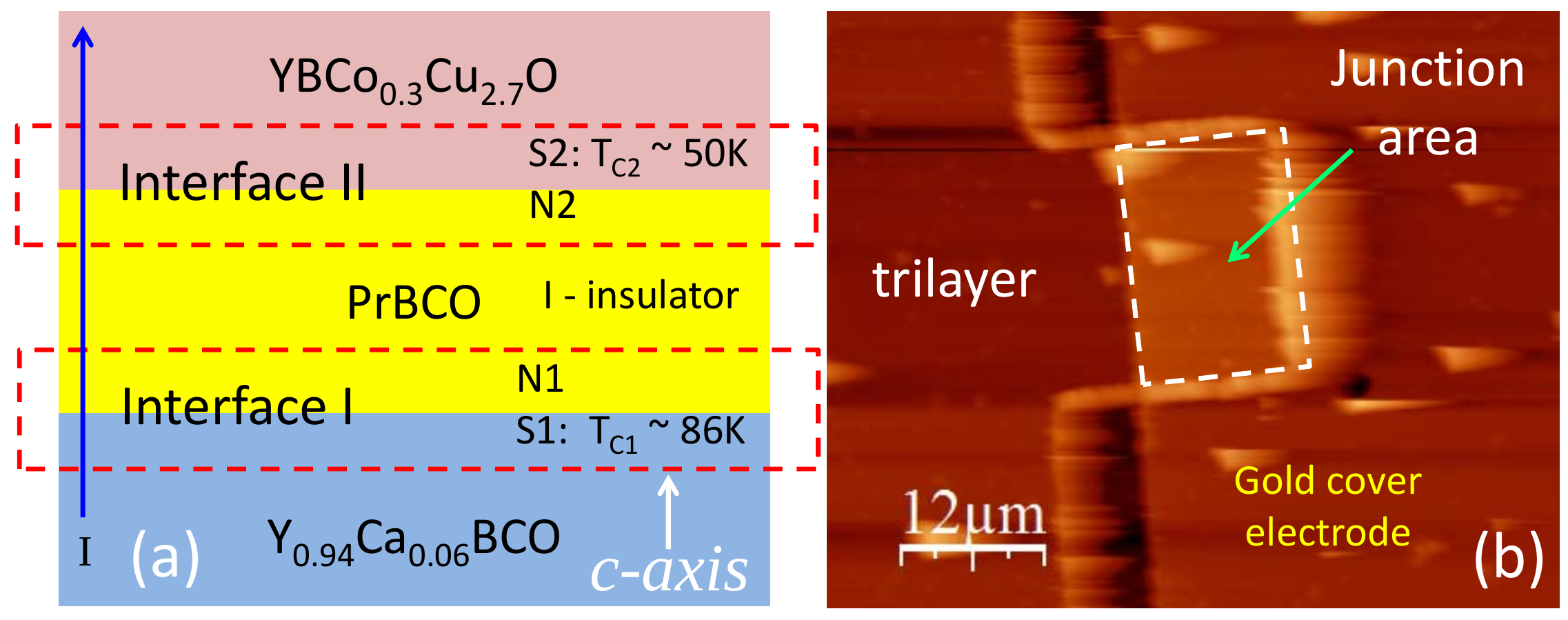}
\vspace{-0mm} \caption{\label{fig:epsart}(a) A schematic model of a trilayer \textit{c-axis} junction cross-section for CJ-4 having 100 nm thick underdoped $YBa_2Co_{0.3}Cu_{2.7}O_y$ on 25 nm insulating $PrBa_2Cu_3O_{7-\delta}$ on 200 nm of overdoped $Y_{0.94}Ca_{0.06}Ba_2Cu_3O_{7-\delta}$. The two interface areas are marked by the dashed rectangles. (b) Atomic force microscope image of a single junction where the gold cover electrode overlaps the trilayer in the marked junction area.
 }
\end{figure}

\section{Results and discussion}

\begin{figure} \hspace{-20mm}
\includegraphics[height=9cm,width=13cm]{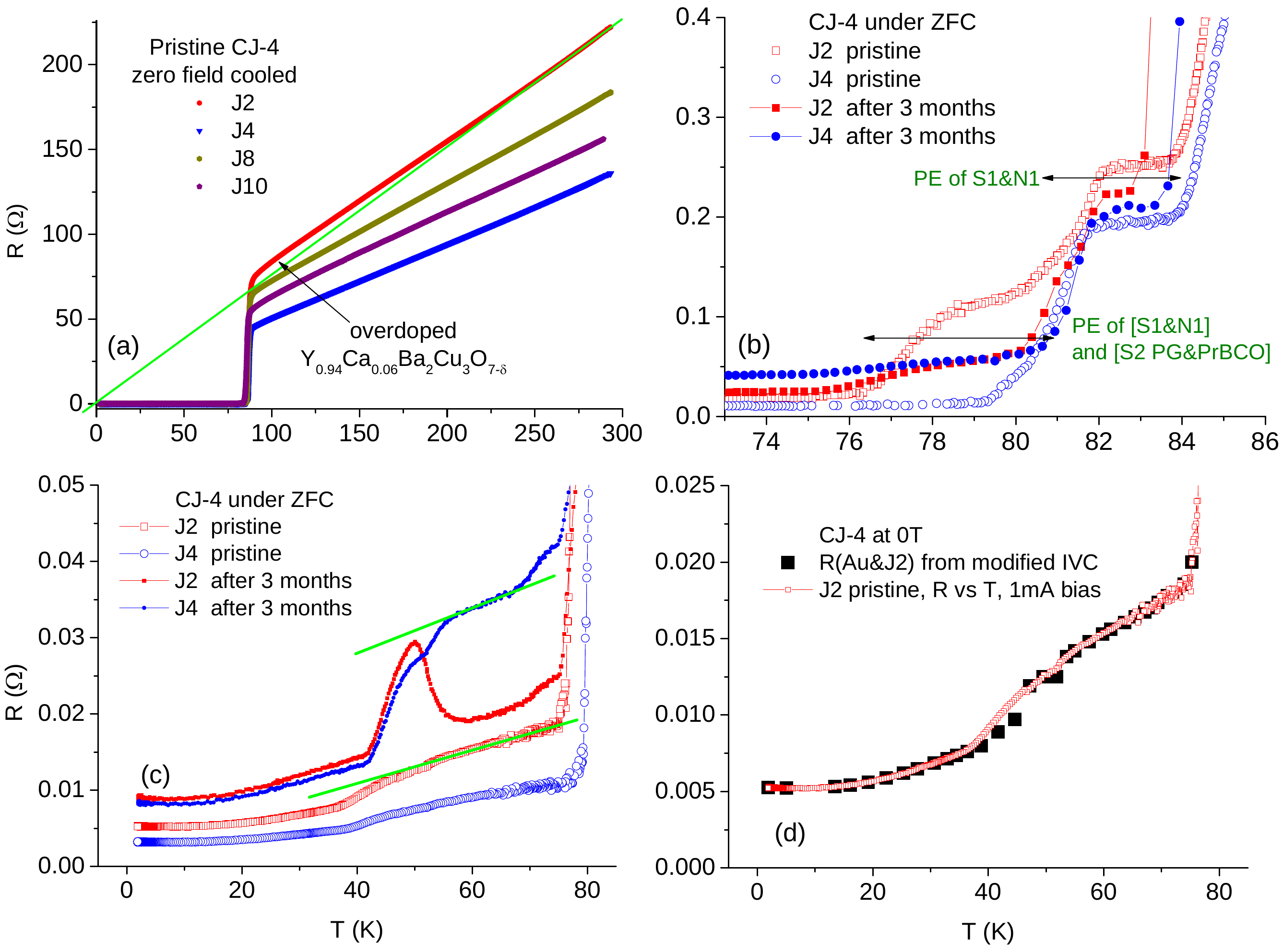}
\vspace{-0mm} \caption{\label{fig:epsart} Zero field cooled (ZFC) resistance versus temperature of junctions on the CJ-4 wafer. (a) Wide temperature range results of four junctions. The different normal resistances are due to different lead lengths. (b) Two proximity transition regimes (steps and double arrows) just below the superconducting transition of $Y_{0.94}Ca_{0.06}Ba_2Cu_3O_{7-\delta}$ at 85-86 K, of two junctions in the pristine and aged states. (c) Zoom in on the the resistive tail down to 2 K, with the YBCoCO transition seen at 50-55 K, and mostly the gold lead resistance below 20 K. (d) A comparison of the R versus T data of the pristine J2 junction measured under 1 mA bias current and the R(Au\&J) extracted from the modified I-V curves as discussed in the text. The YBCoCO transition at about 50 K and its proximity tail down to 40 K are clearly seen here. }

\end{figure}

Zero field cooled resistance versus temperature of a few junctions on the CJ-4 wafer are depicted in Fig. 2. In (a), data for the whole temperature rage measured on the pristine wafer is shown. The different normal state resistance values are due to different lengths of the $Y_{0.94}Ca_{0.06}Ba_2Cu_3O_{7-\delta}$ leads to the junctions. The excess resistance above the straight green line at temperatures higher than the superconducting transition at $T_c\approx 85-86$ K, indicates the overdoped nature of the Ca-doped YBCO electrode. It also shows that the resistances of the gold cover electrode, PrBCO barrier and the YBCoCO layer are negligible in the normal regime. Just below $T_c$, two consecutive proximity transitions are observed as shown in Fig. 2 (b) for the pristine and three months aged junctions. The aging process involved the first measurement cycle and then storage in a desiccator under dry atmosphere for three months. Although $T_c^{onset}$ of the aged sample remained 86 K (not shown), the transition broadened so that $T_c^{offset}$ was about 1 K lower than that of the pristine junctions as seen in (b). The first resistive step (plateau and transition) in the aged sample between 81 and 83-84 K reproduced its pristine behavior quite well, the second step width of about 4 K of J2 remained, but that of J4 increased from about 1 K to 4-5 K. We therefore decided to concentrate in this study on the data of J2, but before that we follow both junctions behavior down to 2 K as seen in Fig. 2 (c). Both the pristine and aged junctions show the superconductive transition of the underdoped YBCoCO electrode with $T_c^{onset}$ at about 50-55 K (see the resistance drops below the green lines). The aged junctions had higher resistances, but while J4 kept a monotonous behavior versus temperature, J2 didn't. J2 showed a peak between 40 and 55 K, in the YBCoCO transition regime. We attribute the increasing part of this peak with decreasing temperature to the insulating PrBCO barrier resistance as in Fig. 1 (a), which is then overcome by the YBCoCO transition to superconductivity that lowers the overall junction resistance again with decreasing temperature. The peak resistance of J2 occurs at about 50 K, which coincides with the kink seen in the J4 data, then they both decrease together due to the proximity effect down to about 42 K, where the resistance almost levels off. This leveling off is due to the serial gold lead resistance which dominates the data below about 20 K.\\

To understand the two steps observed in Fig. 2 (b), we first note that the resistivity $\rho$ of bare thin films of PrBCO is much higher than when the PrBCO layer is incorporated in junctions like we have here. For instance, at 57 K, $\rho_{film}$ ranges between 0.4 to 10$^4$ $\Omega$cm \cite{Fisher}, while in J2 or J4 of CJ-4 it is no more than $\rho_{junction}\approx 0.05$ $\Omega$cm (using the maximum resistance 0.04 $\Omega$ of Fig. 2 (c) and the junction area and PrBCO thickness of Table I). We thus conclude that in junctions the PrBCO layer has a much higher carrier density. Moreover, the decreasing R on lowering the temperature as in Fig. 2 (c), indicates that the two PrBCO layers adjacent to the interfaces in our junctions basically behave like normal metals as depicted by N1 and N2 in Fig. 1 (a). Returning now to the two steps behavior of Fig. 2 (b), we observe that while the first resistance step between 81 and 84 K was common to all the junction on the CJ-4 wafer, the second step between 77 and 80 K wasn't, and appeared fully only in J2, partially in J4, and very weakly in the other junctions. The origin of the first step is simple and results from the conventional proximity effect transition at the Ca-doped YBCO and PrBCO interface marked by "PE of S1\&N1" (referring to the layer numbers in Fig. 1 (a)), and the top double arrow in Fig. 2 (b). The origin of the second step marked by "PE of [S1\&N1] and [S2 PG\&PrBCO]" and the bottom double arrow in Fig. 2 (b) is more subtle (N1 overlaps N2 above  $\sim$70 K). Above $T_c$(YBCoCO), the S2 layer is not superconducting but in the pseudogap (PG) state. Assuming the pre-formed pairs scenario in this regime, one can imagine that uncorrelated pairs penetrate into the PrBCO barrier and making it less resistive. Now since S1\&N1 is already superconducting below 80 K, it induces proximity superconductivity in this less resistive PrBCO layer, leading to the second transition.  We thus conclude that the two steps in Fig. 2 (b) are due to a conventional proximity effect at interface I of Fig. 1 (a), and to an unconventional "second order" one involving preformed pairs in the pseudogap regime in the PrBCO barrier.\\

\begin{figure} \hspace{-20mm}
\includegraphics[height=9cm,width=13cm]{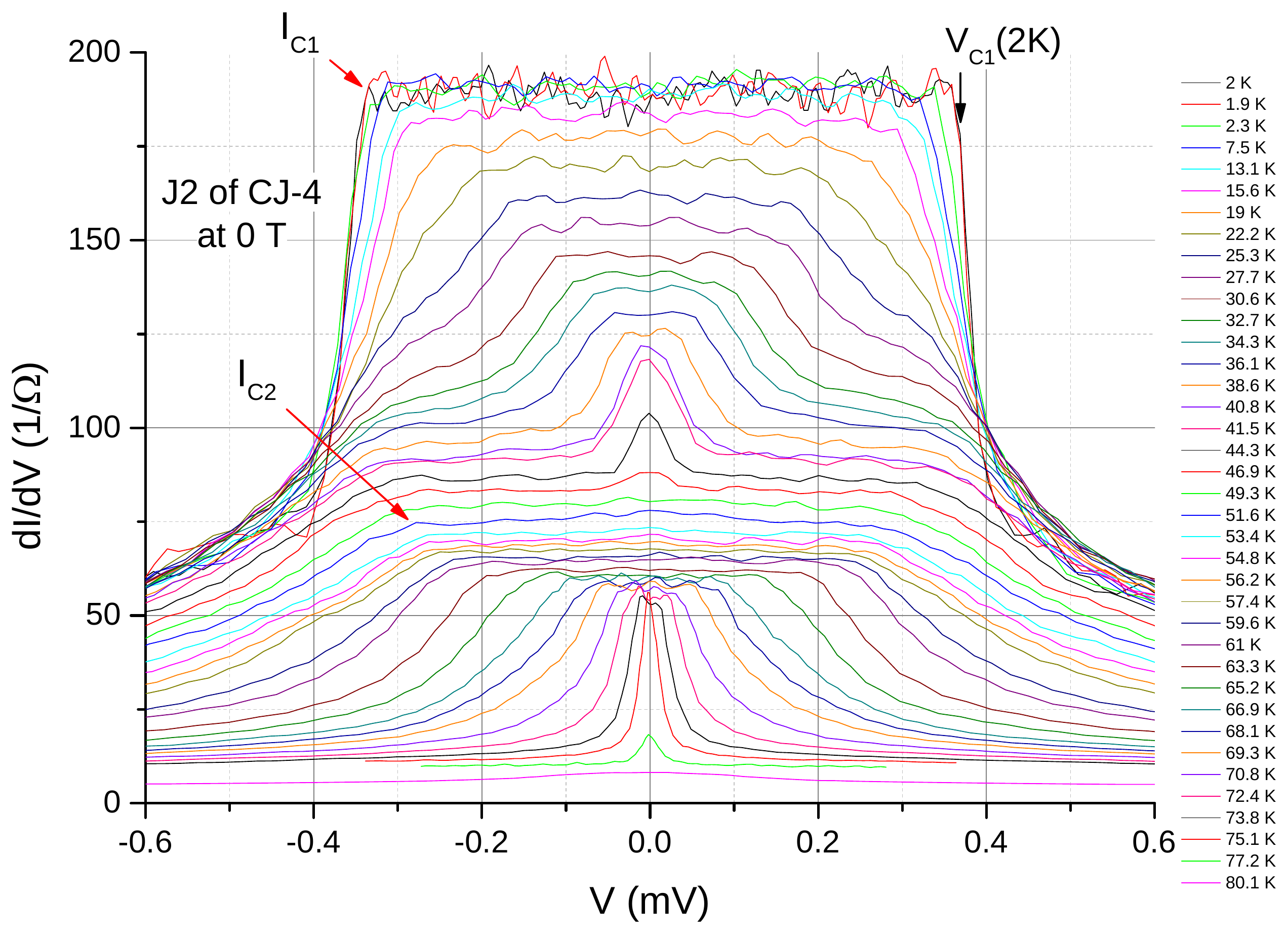}
\vspace{-0mm} \caption{\label{fig:epsart} Conductance spectra of the pristine junction J2 on the CJ-4 wafer at different temperatures and zero field. The flat top-hat shapes indicate reaching the constant serial resistance of the gold and junction R(Au\&J) at each temperature, while the sharp conductance G-drops indicate reaching the corresponding critical currents $I_{c1}$ and $I_{c2}$ as marked by the arrows.     }
\end{figure}

\begin{figure} \hspace{-20mm}
\includegraphics[height=9cm,width=13cm]{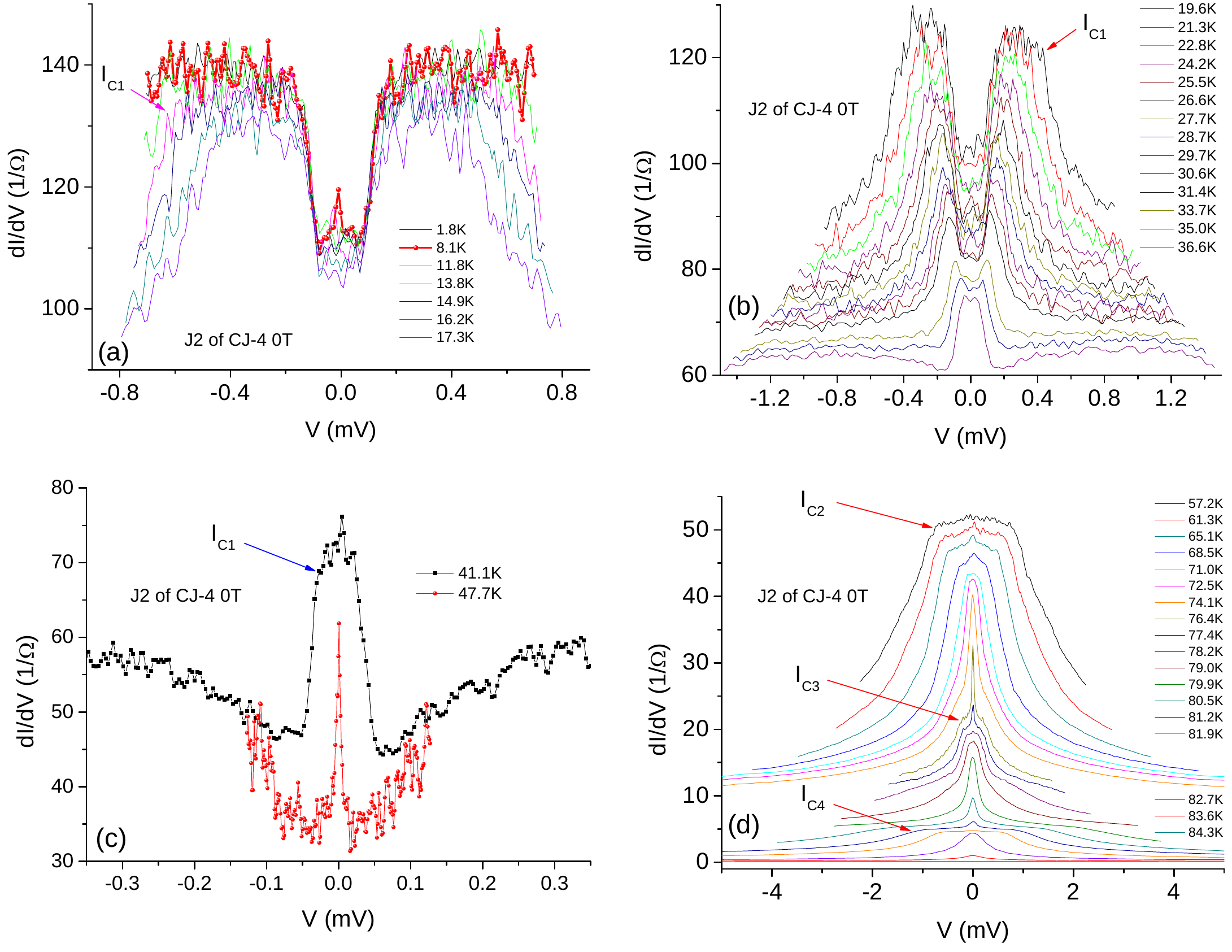}
\vspace{-0mm} \caption{\label{fig:epsart} Conductance spectra of the same J2 junction as in Fig. 3, but after aging for three months in a desiccator. (a) shows that a clear tunneling-like gap developed in the top-hat spectra at low temperatures, with a clear Josephson supercurrent at zero bias. This gap is filled in with increasing temperature as seen in (b), leaving only the top hat at 41.1 K as in (c), which on narrowing down with increasing temperature reveals a very narrow Josephson $I_c$ again at 47.7 K. (d) shows a repeat of this behavior with increasing temperature above $T_c$ of the YBCoCO electrode, i.e. top hat narrowing down, revealing a very narrow Josephson-like $I_c$ at 76.4 K, followed by developing of a new top hat, and so on.
 }
\end{figure}

Conductance spectra of the pristine J2 junction are shown in Fig. 3 at various temperatures from 2 to 80 K and zero magnetic field. At low temperatures, a typical top-hat shape of these spectra is seen which is a result of a constant gold and junction resistance $\rm R(Au\&J)$ up to a critical voltage $\rm V_{c1}$ where the conductance drops sharply, indicating that the critical current $I_{c1}$ in the junction was reached \cite{KorenLee,Kirzhner}. With increasing temperature, these spectra narrow down to a peak around zero bias at 41 to 44 K, where a new top-hat shape develops at higher bias. The later top-hat indicates that another kind of critical current $I_{c2}$ exists in our junctions. Similar conductance spectra were measured after aging of this junction and the detailed results are given in Fig. 4. At low temperatures, the flat part of the top-hat shape remained, but the overall conductance decreased with aging from 200 to 140 $\Omega^{-1}$. In addition, a very small but sharp tunneling-like energy gap $\Delta$ at V$\approx \pm 0.15$ mV developed, with a small Josephson supercurrent at zero bias. Possible origins of this gap could be a superconducting energy gap $\Delta_c$ in the \textit{c-axis} direction, or an induced gap in the PrBCO barrier. On heating up, the top-hat structure narrowed down and the tunneling gap filled up, until a very narrow peak remained at 47.7 K as seen in Fig. 4 (c). Then again, new top hat shapes developed three more times, indicating an overall existence of four supercurrents of different origin $I_{c1}$ to $I_{c4}$ as marked by the arrows in Fig. 4. One observes that apart from the development of the tunneling gap, the results of Figs. 3 and 4. are basically similar. In the following though we shall focus on the aged junction results, which are more detailed and have sharper features, apparently due to the more robust tunneling barrier. \\

\begin{figure} \hspace{-20mm}
\includegraphics[height=9cm,width=13cm]{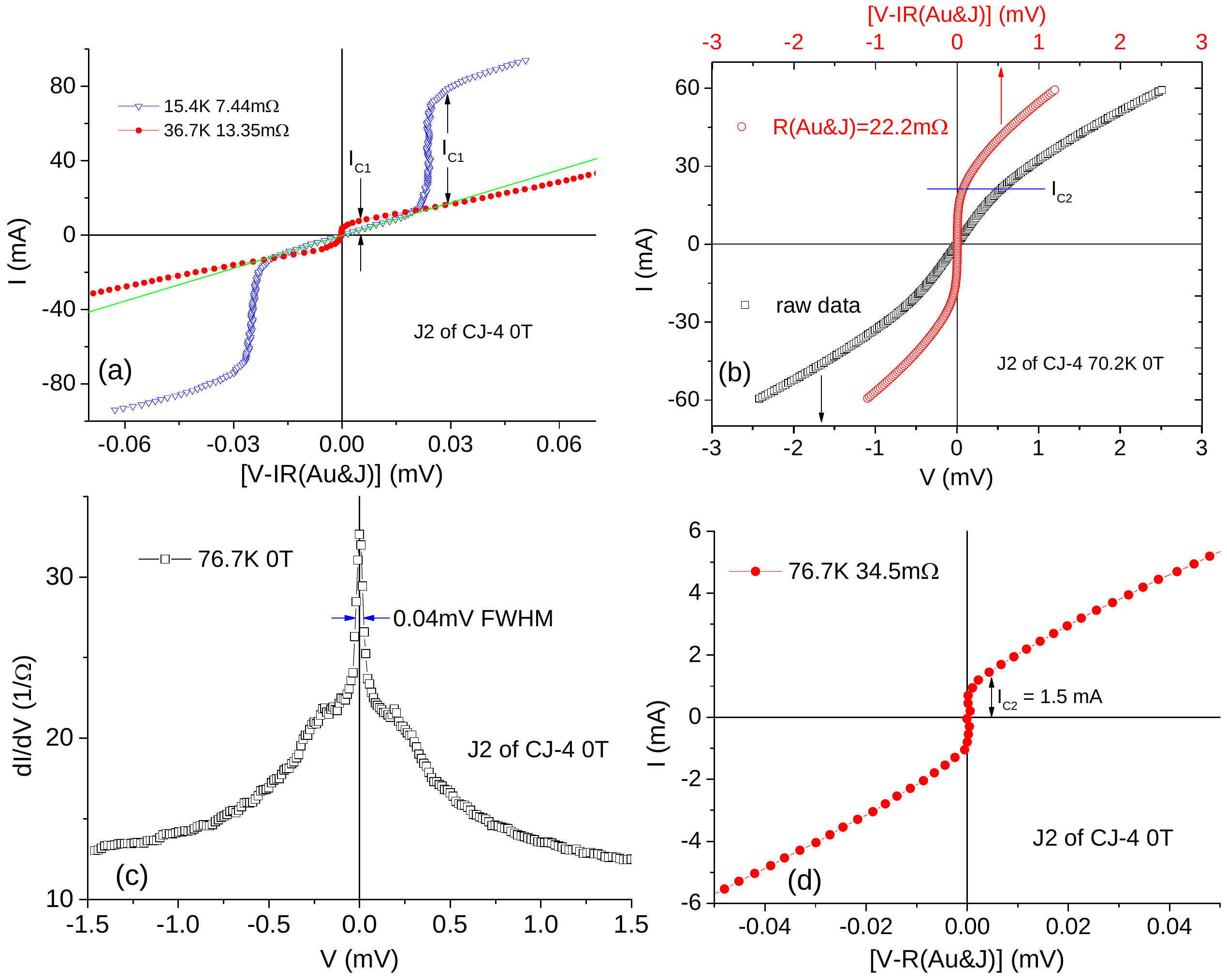}
\vspace{-0mm} \caption{\label{fig:epsart} (a) current versus modified voltage scale of aged  junction J2 of CJ-4 at two temperatures. The corresponding $I_{c1}$ values using a 5 $\mu$V criterion are marked with arrows. (b) shows a raw data I-V curve at 70.2 K versus V (bottom scale), and a modified curve of the same data after subtracting the the voltage drop on the gold lead and junction resistance from the measured V (top scale). Conductance spectrum at 76.7 K with a very narrow peak is depicted in (c), with the corresponding modified IVC in (d) from which a clear $I_{c2}$ value can be obtained.
}
\end{figure}

To determine the different critical currents we used the current versus voltage I-V data that was measured simultaneously with the conductance spectra. Since our junctions always had a serial resistance R, we had to subtract the voltage drop IR on it from the measured voltage V  in order to have I-V curves (IVC) with finite current at zero voltage (generally, IVC with a vertical segment at V=0), from which $\rm I_c$ could be determined. Two factors contributed to the serial resistance R, one was the resistive gold lead to the junction, and the other the resistance of the junction itself. Thus we mark it as R=R(Au\&J). Fig. 5 depicts a few representative such I-V curves ("modified" I-V curves) of the J2 junction of CJ-4 under zero field, where the current is plotted versus V-IR(Au\&J), and the critical currents $\rm I_c$ is determined by a $5\,\mu V$ voltage criterion. Fig. 5 (a) shows two low temperature results. The curve at 15.4 K depicts a typical quasiparticles tunneling IVC with a very small junction resistance background (here R=R(Au) only), but without a Josephson supercurrent at zero bias, due to the current rather than voltage bias used in our measurements. At 36.7 K, the tunneling behavior disappears (see also Fig. 4 (b)), and a more conventional IVC is seen (here R=R(Au\&J) again). In (b), a raw data IVC at 70.2 K together with the modified one are shown. The kink in the raw data curve is much clearer in the modified curve, the rounding is due to flux flow at the high temperature, and this is the reason for choosing a relatively high voltage criterion (5 rather than 1 $\mu$V) for the determination of $\rm I_c$. Fig. 5 (c) and (d) depict conductance spectra and the corresponding modified IVC at 76.7 K. The very narrow peak of the conductance near zero bias, leads to a very sharp modified IVC with very little rounding, unlike the data in (b) though the data in (d) is taken at a higher temperature. A similarly sharp modified IVC is obtained also for the data at 47.7 K (not shown), where a very narrow zero bias conductance peak has also been observed (Fig. 4 (c)).
As a reliability test of our procedure of using the modified IVC by subtracting IR(Au\&J) from the measured voltage V, we plotted in Fig. 2 (d) the resulting R(Au\&J) (a few values of which are given in Fig. 5) versus T, together with the directly measured R versus T data of the pristine J2 junction measured under 1 mA current bias. The nice overlap of the two sets of data, attests to the reliability of our method.\\

\begin{figure} \hspace{-20mm}
\includegraphics[height=9cm,width=13cm]{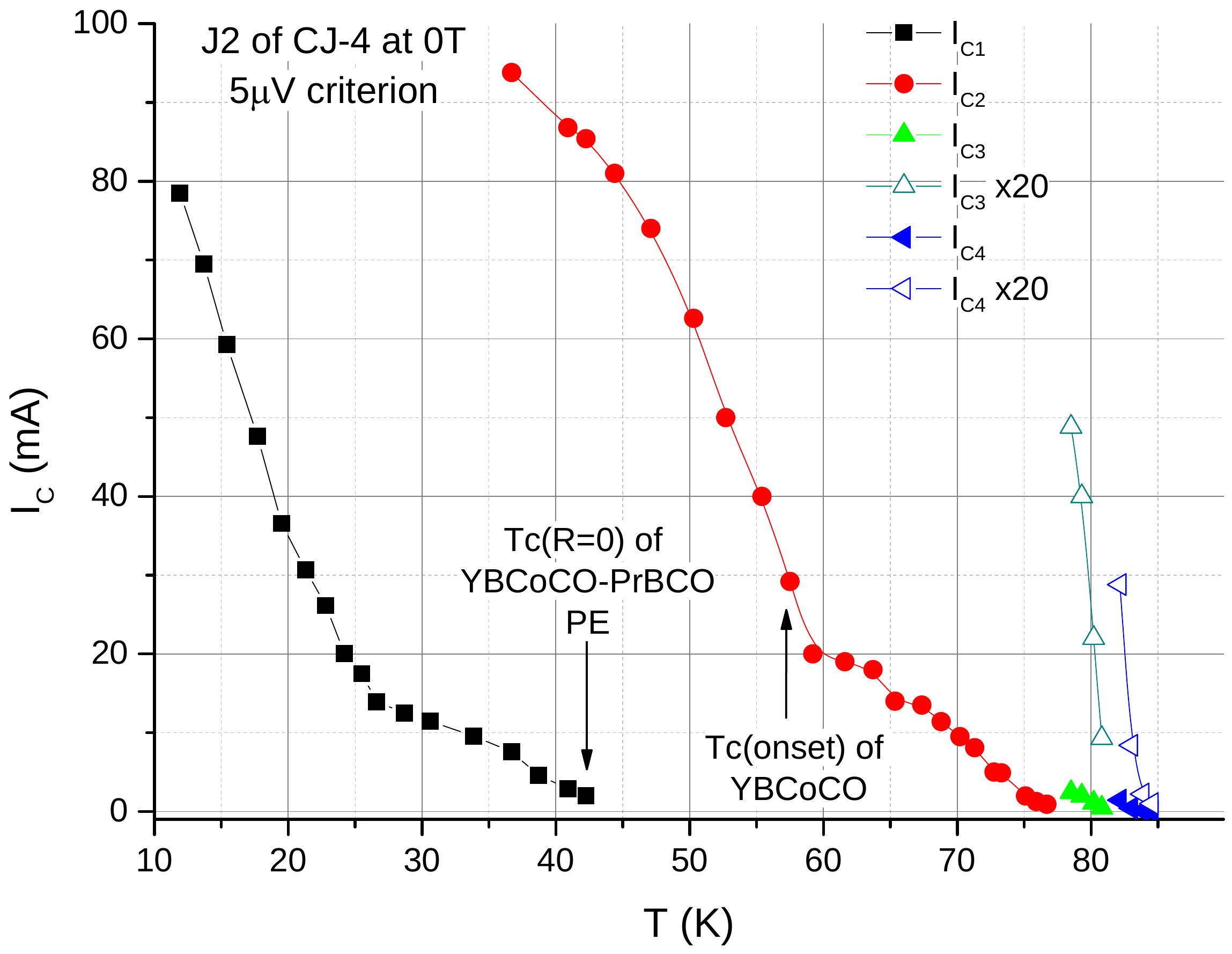}
\vspace{-0mm} \caption{\label{fig:epsart}
The four critical currents $I_{ci}$ deduced from the data of Fig. 4 after plotting the corresponding modified IVC and using the 5 $\mu$V criterion as shown for a few cases in Fig. 5 (a), (b) and (d). $I_{c1}$ is the coherent supercurrent when both electrodes are superconducting, $I_{c2}$ is the fluctuating pairs current, at temperatures larger than about 40 K which is the superconducting transition of the proximity layer N2 of interface-II of Fig. 1 (a).  $I_{c3}$ and $I_{c4}$ are attributed to the two proximity transitions of the two steps in Fig. 2 (b), see text.           }
\end{figure}

The four different critical currents obtained in this study under zero field are plotted versus temperature in Fig. 6. The low temperature $I_{c1}$ is the coherent supercurrent where both electrodes are in the superconducting state. The reason it starts rising just below 42 K is that this is the temperature at which the YBCoCO-PrBCO proximity transition ends, see Fig. 2 (c). In addition, the kink at 26 K seems to be due to a vanishing junction resistance, leaving only the serial resistance of the gold lead (R=R(Au) only), in full agreement with Figs. 2 (c) and 5 (a). $I_{c2}$ is basically a fluctuation pairs current in the pseudogap regime of YBCoCO. It persists up to 77 K which is about 20 K above the onset of the YBCoCO transition temperature to superconductivity as seen in Fig. 2 (d).  This $\rm T_c^{onset}(YBCoCO)\approx$ 57 K \cite{KorenPolturak}, is also clearly seen in Fig. 6 by the sharp increase of $I_{c2}$ on cooling down below this temperature. Due to a current source limit of 100 mA, and heating effects at this high bias current, we could not measure $I_{c2}$ below about 36 K. It is unclear if $I_{c2}$ at low temperatures (below about 50 K) has a coherent component or not. The last two pair currents $I_{c3}$ and $I_{c4}$ above 77 K are attributed to the two proximity regions elaborated on previously, and marked by the two double arrows in Fig. 2 (b).\\

\begin{figure} \hspace{-20mm}
\includegraphics[height=9cm,width=13cm]{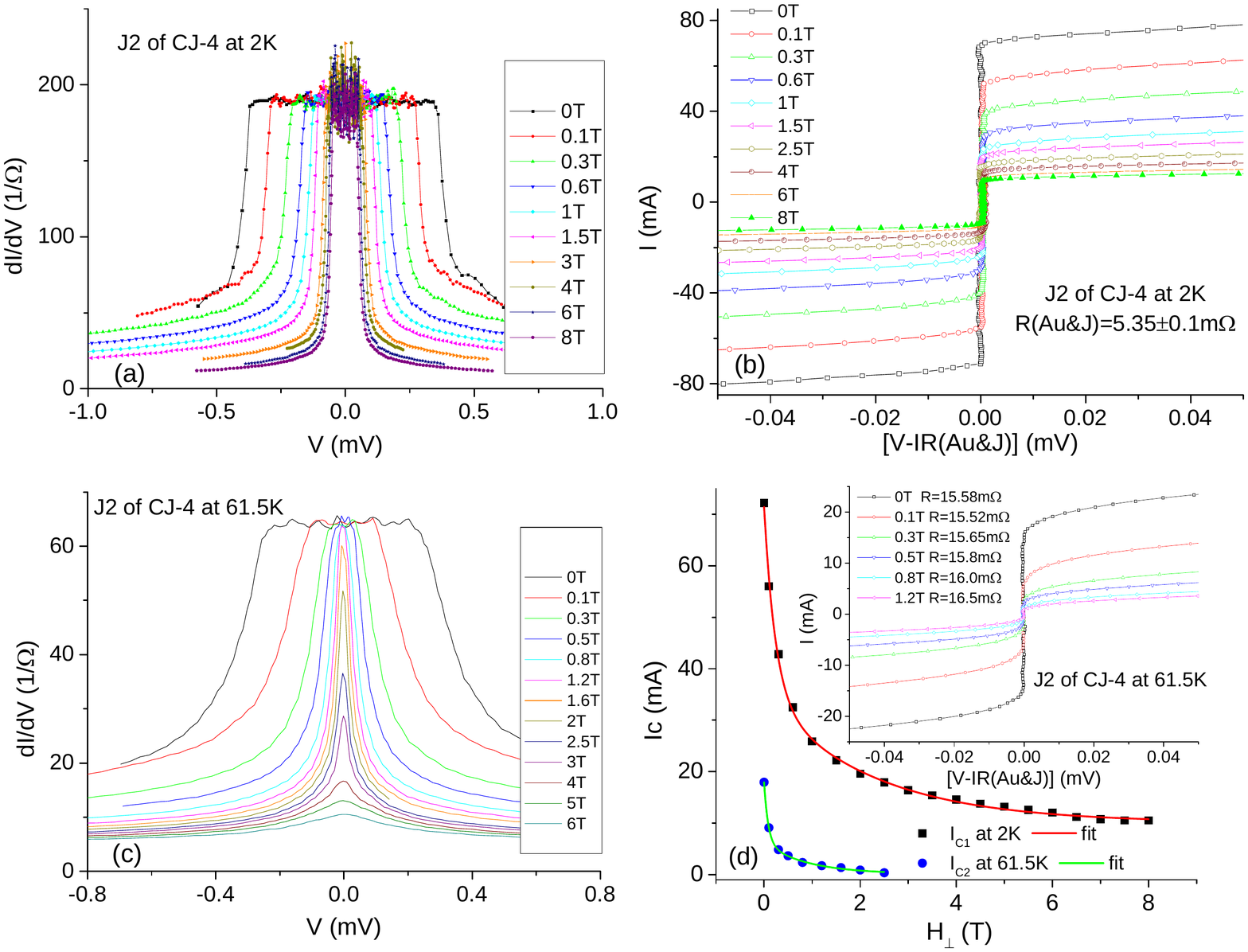}
\vspace{-0mm} \caption{\label{fig:epsart} Conductance spectra of the pristine J2 junction of the CJ-4 wafer at 2 K (a), and 61.5 K (c), under various magnetic fields, and the corresponding modified IVC in (b), and the inset to (d), respectively. The main panel in (d) shows $I_{c1}$ at 2K and $I_{c2}$ at 61.5 K (determined by the 5 $\mu$V criterion) versus field, together with double-exponential decay fits to this data. The field decay parameters of the fits are $H_S$=0.24 and $H_L$=2.2 T at 2 K, and $H_S$=0.09 and $H_L$=0.9 T at 61.5 K.
}
\end{figure}

Next, we go back to the pristine J2 junction of the CJ-4 wafer and present its behavior under magnetic fields normal to the a-b planes, at temperatures below and above $\rm T_c$(YBCoCO). Fig. 7 (a) shows conductance spectra at 2 K under various fields, while in (b) the corresponding modified IVCs are shown. Similarly, conductance spectra and modified IVCs at 61.5 K, which is a few degrees above $\rm T_c(YBCoCO)\approx 55$ K, are shown in (c) and the inset to (d), respectively. The critical currents $I_{c1}$ at 2K and $I_{c2}$ at 61.5 K were found by applying the 5 $\mu$V criterion to the modified IVC, and plotted in the main panel of (d) versus field. The curves in (d) are phenomenological double-exponential decay fits of $I_c=I_0+I_S\, exp(-H/H_S)+ I_L\, exp(-H/H_L)$ to the data, where $I_0$, $I_S$ and $I_L$ are constants, and $H_S$ and $H_L$ are the small and large field decay parameters. We note that fits to a single exponential decay didn't fit the data. Clearly, the field suppresses superconductivity in the junction electrodes at both temperatures. We attribute the two different field decay parameters at each temperature to the two different electrodes of the junction. The larger field decay parameters ($H_L$=2.2 and 0.9 T) are attributed to the more robust higher $T_c$ overdoped $Y_{0.94}Ca_{0.06}Ba_2Cu_3O_{7-\delta}$ electrode, while the smaller field decay parameters ($H_S$=0.24 and 0.09 T) correspond to the lower $T_c$ underdoped $YBa_2Cu_{2.7}Co_{0.3}O_y$ electrode. One can see that the ratio $H_L/H_S$ at each temperature (9.2 at 2 K and 10 at 61.5 K) are quite similar. On the other hand, the constants $I_0$ are very different, 10.1 mA at 2 K and 0.14 mA at 61.5 K. Thus if the $I_0$ contributions to $I_c$ are subtracted, the result is a similar contribution of the two field decay components at each of the two temperatures (same curve shapes). The ratios $H_S$(2 K)/$H_S$(61.5 K)=2.7 and $H_L$(2 K)/$H_L$(61.5 K)=2.5 are also comparable. Therefore, disregarding whether we compare the large or small field decay parameters at the two temperatures, there is a clear evidence that at 61.5 K in the pseudogap regime of the YBCoCO electrode, both $H_S$ and $H_L$ are much smaller than at 2 K.\\

\section{Conclusions}

An unconventional, second order proximity effect was observed in S1/I/S2 junction, in which a standard proximity region at the interface of a strong S1 superconductor and an adjacent N1 part of a PrBCO barrier, induces a second proximity effect in this barrier which contains injected preformed pairs from a weaker S2 superconducting electrode in its pseudogap regime. Another result was that four different pair currents were observed as a function of temperature, which were attributed to coherent supercurrents at low temperatures, to fluctuating pairs current above $\rm T_c(S2)$, and to the two proximity regimes just below $\rm T_c(S1)$. Finally, the pairs current was suppressed under magnetic fields, with two distinct decay parameters that originated in the two different electrodes, with a significant suppression observed in the pseudogap regime of S2.\\

{\em Acknowledgments:}
The author is grateful to Patrick. A. Lee and Assa Auerbach for useful discussions.\\

\bibliography{AndDepBib.bib}

\bibliography{apssamp}

\end{document}